\documentclass[aps,prl,twocolumn,groupedaddress,showpacs,floatfix]{revtex4}
\usepackage{graphicx}
\usepackage{amsmath}
\usepackage{color}
\begin{document}
\pagenumbering{arabic}
\pagestyle{plain}
\title{Observation of Orbital Ordering and Origin of the Nematic Order in FeSe}
\author{R. X. Cao,$^{1,2}$ Jian Hu,$^{1}$ Jun Dong,$^{1}$ J. B. Zhang,$^{1}$ X. S. Ye,$^{1}$ Y. F. Xu,$^{1}$ D. A. Chareev,$^{3,4,5}$ A. N. Vasiliev,$^{6,7,8}$ Bing Wu,$^{9}$ X. H. Zeng $^{1,\ast}$, Q. L. Wang, $^{10,\ast}$ and Guoqing Wu $^{1,\ast}$}
\affiliation{$^{1}$College of Physics Science and Technology, Yangzhou University, Yangzhou, Jiangsu 225002, China}
\affiliation{$^{2}$National Laboratory of Solid State Microstructures, Nanjing University, Nanjing 210093, China}
\affiliation{$^{3}$Institute of Experimental Mineralogy, Russian Academy of Sciences, 142432, Chernogolovka, Moscow District, Russia}
\affiliation{$^{4}$Institute of Physics and Technology, Ural Federal University, Mira st. 19, Ekaterinburg, 620002, Russia}
\affiliation{$^{5}$Kazan Federal University, 18 Kremlyovskaya Str., Kazan, 420008, Russia}
\affiliation{$^{6}$Low Temperature Physics and Superconductivity Department, Lomonosov Moscow State University, Moscow 119991, Russia}
\affiliation{$^{7}$National University of Science and Technology ``MISiS", Moscow 119049, Russia}
\affiliation{$^{8}$National Research South Ural State University, Chelyabinsk 454080, Russia}
\affiliation{$^{9}$Department of Math and Computer Science, Fayetteville State University, Fayetteville, NC 28301, USA}
\affiliation{$^{10}$Institute of Electrical Engineering, Chinese Academy of Sciences, Beijing 100190, China}
%
%
%
\begin{abstract}

    In iron-based superconductors the interactions driving the nematic order that breaks the lattice four-fold rotational symmetry in the iron plane may also facilitate the Cooper pairing, but experimental determination of these interactions is challenging because the temperatures of the nematic order and the order of other electronic phases appear to match each other or to be close to each other. Here we performed field-dependent $^{77}$Se-nuclear magnetic resonance (NMR) measurements on single crystals of iron-based superconductor FeSe, with magnetic field $B_{0}$ up to 16 T. The $^{77}$Se-NMR spectra and Knight shift split when the direction of $B_{0}$ is away from the direction perpendicular to the iron planes (i.e., $B_{0}$ $\parallel$ $c$) upon cooling in temperature, with a significant change in the distribution and magnitude of the internal magnetic field at the $^{77}$Se nucleus, but these do not happen when $B_{0}$ is perpendicular to the iron planes, thus demonstrating that there is an orbital ordering. Moreover, stripe-type antiferromagnetism is absent, while giant antiferromagnetic spin fluctuations measured by the NMR spin-lattice relaxation gradually developed starting at $\sim$ 40 K, which is far below the nematic order temperature $T_{\text{nem}}$ = 89 K. These results provide direct evidence of orbital-driven nematic order in FeSe.

\end{abstract}
%
\thanks{Corresponding authors: wugq@yzu.edu.cn (Guoqing Wu), xhzeng@yzu.edu.cn (X. H. Zeng), and qiuliang@mail.iee.ac.cn(Q. L. Wang)}
\maketitle
\section*{1. Introduction}
    The interactions between structure, magnetism and superconductivity in Fe-based superconductors have been of wide interests \cite{Paglione-2010, Fernandes-2010, Tanatar-2016}. The experimental determination of these interactions is challenging due to the occurrence of nematic order often at or near the temperature of a stripe-type long-range antiferromagnetic (AFM) order \cite{Kontani-2011, Fernandes-2013, Nakai-2013, Bohmer-2013, Fernandes-2014}. Similar to the stripe-type AFM order, the nematic order also breaks the lattice four-fold ($C_{4}$) rotational symmetry of a high-temperature phase, as evidenced by a tetragonal-to-orthorhombic structural phase transition at $T_{\text{s}}$ \cite{Bohmer-2013, Fernandes-2014, Chu-2010, Konstantinova-2019, Chen-2019}. On the other hand, the nematic order is directly linked to the superconducting state because nematic instability is a characteristic feature of the normal state upon which at lower temperatures the superconductivity emerges \cite{Paglione-2010, Fernandes-2014, Kang-2018, Liu-2018}, and thus nematicity is deemed a precursor of superconductivity in unconventional superconductors including the cuprates. It is generally believed that the nematic order is electronic and the structural phase transition is the consequence of the nematic order, since the lattice distortion is much smaller than the observed anisotropy of the in-plane resistivity in the nematic phase \cite{Chu-2010, Tanatar-2010}. However, it remains highly controversial regarding the origin of the nematic order whether it is driven by spin order \cite{Fang-2008, Hu-2012}, AFM spin fluctuations \cite{Fang-2008, Hu-2012, Wang-2016}, and/or orbital order \cite{Lv-2009, Lee-2009, Kruger-2009, Daghofer-2010, Chen-2010, Baek-2015, Bohmer-2015}.

    In Fe-based superconductors, FeSe has the simplest crystal structure, while it has representative properties as other Fe-based superconductors, thus it has been intensively studied \cite{Hsu-2008, Buchner-2009, Stewart-2011}. FeSe undergoes a tetragonal-to-orthorhombic structural phase transition at $T_{\text{s}} \sim$ 90 K with an electronic nematic order simultaneously ($T_{\text{nem}}$ = $T_{\text{s}}$) \cite{McQueen-2009, Nakayama-2014, Shimojima-2014, Tanatar-2016}. The orbital ordering was also found at $T_{\text{nem}}$ via angle-resolved photoemission spectroscopy (ARPES) \cite{Nakayama-2014, Watson-2015, Zhang-2015}, whereas AFM order was absent at ambient pressure \cite{McQueen-2009, Bendele-2010, Mizuguchi-2010}, and thus possible orbital order driven nematicity was proposed \cite{Nakayama-2014, Watson-2015, Zhang-2015, Tanatar-2016}.

    However, recent findings show that stripe-type AFM order emerges under high pressure, and the AFM ordering temperature increases with high pressure \cite{Imai-2009, Bendele-2010, Sun-2016, Kothapalli-2016, Wang-2016}. These findings make the origin of the electronic nematic order more elusive. Even though various experimental techniques have been used for the study, most research work reported was focused on the doping and high pressure effects on the properties of FeSe. A systematical investigation of the effect of applied magnetic field on the properties of FeSe is still lacking.

    Here we present for the first time field-dependent $^{77}$Se-NMR measurements systematically on high-quality single crystals of FeSe with applied magnetic field $B_{0}$ up to 16 T and temperature down to 1.5 K. We observed orbital ordering which is demonstrated by the splitting of the NMR spectrum and Knight shift. As measured by the $^{77}$Se-NMR spin-lattice relaxation, giant AFM spin fluctuations gradually develop starting at $\sim$ 40 K, which is far below the nematic order temperature $T_{\text{nem}}$ = $T_{\text{s}}$ = 89 K. These discoveries provide direct evidence of orbital-driven nematic order in FeSe. They also shed light on the important role of the nematic order on the superconductivity of Fe-based superconductors.
\section*{2. Experimental Section}
     Single crystals of FeSe were grown in evacuated quartz ampoules using the AlCl$_{3}$/KCl flux technique with a temperature gradient of 5$^{o}$C/cm along the ampoule length. The temperatures of the hot and cold ends used for the growth were 427$^{o}$C and 380$^{o}$C, respectively. X-ray diffraction verified that the crystals have a high-purity single phase with a tetragonal crystal structure at room temperature, where the lattice $c$-axis is perpendicular to the Fe-planes ($ab$-plane). A SQUID magnetometer was used to measure the DC magnetic susceptibility $\chi(T)$. The samples used for our NMR measurements have a typical size of 3.3 $\times$ 2.7 $\times$ 0.1 mm$^{3}$.

     The NMR coil used for the measurements was made from 50 $\mu$m diameter silver wire wound with $\sim$ 18 turns and attached to a goniometer on an NMR probe by epoxy. A single-crystal FeSe sample was put inside the coil so that the sample rotation axis is in the lattice $ab$-plane and perpendicular to the applied field $B_{0}$. A commercial network analyzer was used for the observation of the tuning and matching of the resonant circuit located at the bottom of the NMR probe. The NMR spectra and spin-lattice relaxation time $T_{1}$ were measured with an inversion-recovery method, where a $\pi$ pulse is first applied to invert the nuclear magnetization $M_{0}$ to the -$z$ axis, and then after a delay time $t$, a $\pi$/2 pulse is applied to measure the recovering magnetization $M$($t$) component along the $z$ axis, which gives $T_{1}$ as a function of time $t$ as $M$($t$) / $M_{0}$ = 1 $-$ 2exp( $-$ $t$ / $T_{1}$) \cite{Fukushima-1981}.
\section*{3. Results and Discussion}
     Figure 1(a) shows the typical $^{77}$Se-NMR spectra at $B_{0}$ = 12 T and temperature $T$ = 40 K (below $T_{\text{nem}}$), by the variation of the angle $\theta$ between $B_{0}$ and the lattice $c$-axis of FeSe. The spectrum splits  into two peaks (P$_{1}$ and P$_{2}$) which is observed when $B_{0}$ is applied $\sim$ 25$^{o}$ from the $c$-axis. The splits reach the maximum when $B_{0}$ $\parallel$ $a\&b$, indicating the largest anisotropy of the internal field at the Se-sites in the $a\&b$-plane.

     Figure 1(b) exhibits the temperature ($T$) dependence of the $^{77}$Se-NMR spectrum linewidth ($\Delta f$) at $B_{0}$ $\parallel$ $c$ and $\parallel$ $a\&b$, from the measurements of the temperature-dependent $^{77}$Se-NMR spectra at $B_{0}$ = 12 T for both field directions. It shows a significant change at $B_{0}$ $\parallel$ $a\&b$, but not at $B_{0}$ $\parallel$ $c$.

     Noticeably, the $^{77}$Se-NMR spectra are fully magnetic with no electron charge or quadrupolar contributions because $^{77}$Se is a spin $I$ = 1/2 nucleus (which has no quadrupole moment). The spectral splitting was only observed when temperature $T$ is less than $T_{\text{nem}}$ = $T_{\text{s}}$ = 89 K, but not for $B_{0}$ $\parallel$ $c$ (the reason to be revealed later).

     Thus, undoubtedly the spectrum split is the result of a structure symmetry break in the $ab$-plane due to the tetragonal-to-orthorhombic structure phase transition (structurally $a$ and $b$ are not equal any more), which is known as the consequence of the electronic nematic order in the Fe-planes \cite{Baek-2015, Bohmer-2015}.

     Moreover, with the nematic order the spectrum splits also reflect a significant change in spacial field distribution ($\Delta$$B_{\text{FWHM}}$) and also a change in the value of the internal field ($B'$) at the Se-sites. Here $\Delta$$B_{\text{FWHM}}$ = $\Delta f$/$^{77}\gamma_{I}$, where $^{77}\gamma_{I}$ = 8.131 MHz/T is the gyromagnetic ratio of the $^{77}$Se nucleus, and $B'$ = ($\nu$ $-$ $\nu_{L})$/$^{77}\gamma_{I}$, where $\nu$ is the NMR frequency and $\nu_{L}$ is the Larmor frequency ($\nu_{L}$ = $^{77}\gamma_{I}$$B_{0}$).
\begin{figure}
\includegraphics[scale= 0.23]{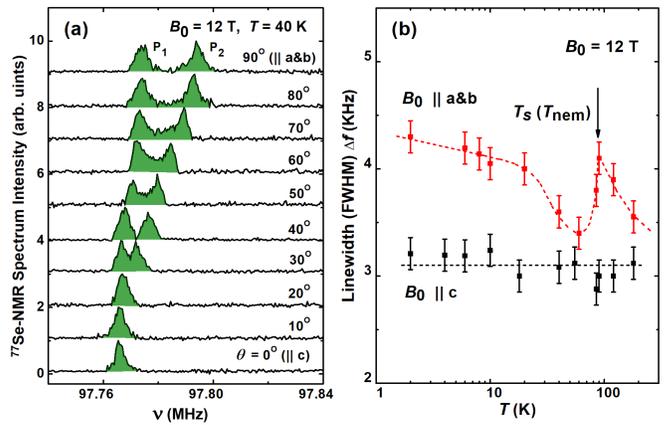}
\caption{(a) $^{77}$Se-NMR spectra of FeSe measured at $B_{0}$ = 12 T and $T$ = 40 K as a function of angle $\theta$, where $\theta$ is the angle between $B_{0}$ and the lattice $c$-axis. (b) $T$-dependence of the $^{77}$Se-NMR linewidth (FWHM) $\Delta f$ at $B_{0}$ $\parallel$ $a\&b$ and $B_{0}$ $\parallel$ $c$. The dashed lines are guides to the eyes. \label{fig1}}
\end{figure}
\begin{figure*}
\includegraphics[scale= 0.35]{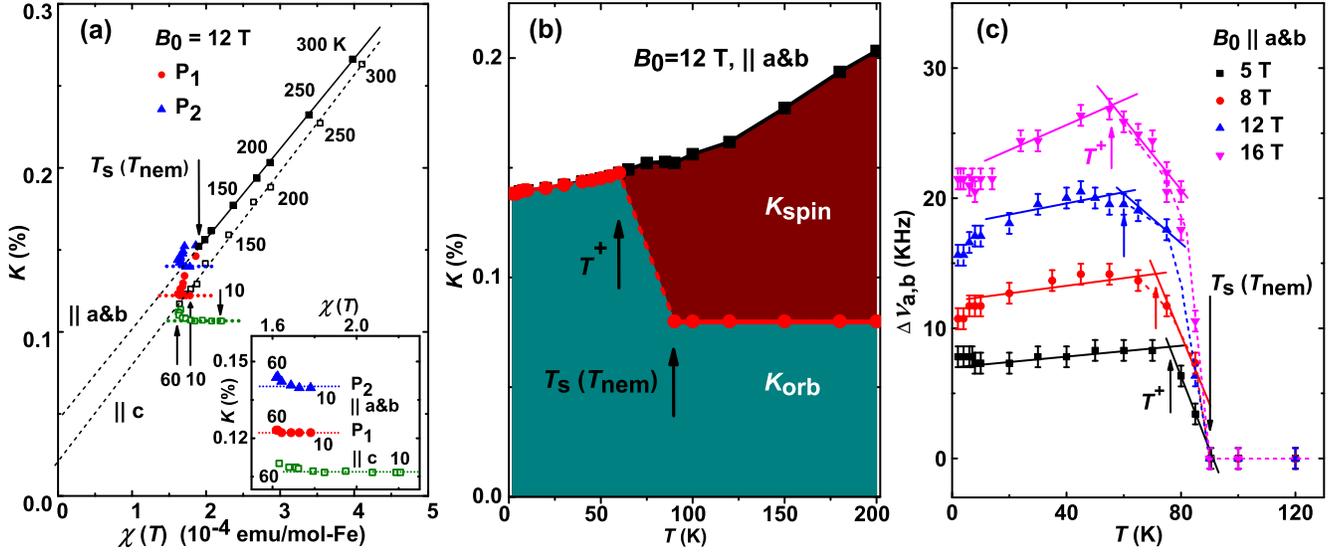}
\caption{(a) Knight shift $K(T)$ versus magnetic susceptibility $\chi(T)$ plot, where the straight lines are for the slopes above $T_{\text{nem}}$. The inset of (a) is an enlargement for the data at $T$ = 60 - 10 K (Note: the critical temperature of superconductivity $T_{c}$ $<$ 10 K). (b) $^{77}$Se-NMR Knight shift $K(T)$ versus $T$ including the contributions of the spin Knight shift $K_{\text{spin}}$ and orbital Knight shift $K_{\text{orb}}$ when $B_{0}$ $\parallel$ $a\&b$-plane. (c) Field-dependence of the difference of the in-plane NMR spectrum peak frequency ($\Delta \nu_{a, b}$) upon cooling in temperature. The dashed lines in (c) are guides to the eyes and the solid lines in (c) are the fit to determine $T^{+}$.  \label{fig2}}
\end{figure*}

     For example, above $T_{\text{nem}}$ the linewidth $\Delta f$ = 3.5 kHz at $T$ = 200 K. Upon cooling in temperature, it goes up and it reaches a maximum of 4.2 kHz at $T$ = $T_{\text{nem}}$ [Figure 1(b)], followed by a complete separation of the two NMR spectrum peaks, and $\Delta f$ = 3.5 kHz at $T$ $\sim$ 60 K again. Then $\Delta f$ increases upon further cooling. While the spectrum linewidth $\Delta f$ (FWHM) at $B_{0}$ $\parallel$ $c$ keeps no change down to low $T$ [Figure 1(b)].

     At the meantime, the value of the internal field $B'$ at the Se-sites has a change $\Delta B'$ = $\pm$ 12.0 G (a Knight shift change of $\pm$ 0.010$\%$) from the average value ($\overline{B'}$) of the internal field $\overline{B'}$ = 160 G (or an average Knight shift 0.133$\%$) (not shown here), i.e., the change of the value of internal field $\Delta B'$ reaches $\pm$ 7.5$\%$ from the average value of the internal field $\overline{B'}$ in the Fe-planes.

     The Knight shift $K$ is defined by $K$ = ($\nu$ $-$ $\nu_{L}$)/$\nu_{L}$ as a tradition, and it should be field independent. That the values of $K(T)$ at $B_{0}$ $\parallel$ $a\&b$ are apparently larger than those at $B_{0}$ $\parallel$ $c$ at $T$ $<$ $T_{\text{nem}}$ indicates an anisotropic hyperfine coupling.

     In general, the Knight shift $K$ is given by \cite{Slichter-1989, Kotegawa-2008}:
\begin{equation}
    K = K_{\text{spin}} + K_{\text{orb}},
\end{equation}
      where spin Knight shift $K_{\text{spin}}$ = [$A_{\text{spin}}$/$N_{A}$$\mu_{B}$]$\chi_{\text{spin}}$, and orbital Knight shift $K_{\text{orb}}$ = [$A_{\text{orb}}$/$N_{A}$$\mu_{B}$]$\chi_{\text{orb}}$. Here $\chi_{\text{spin}}$ and $\chi_{\text{orb}}$ are the electron spin and orbital susceptibility, respectively. $A_{\text{spin}}$ and $A_{\text{orb}}$ are the hyperfine coupling constants between the studied nucleus and the electron spins and the electron orbitals, respectively. $N_{A}$ is the Avogadro's number and $\mu_{B}$ is the Bohr magneton. Likewise, the magnetic susceptibility $\chi$ is the sum of the contributions from core diamagnetic susceptibility ($\chi_{\text{dia}}$), orbital (van Vleck) paramagnetic susceptibility ($\chi_{\text{orb}}$) and Pauli spin paramagnetic susceptibility ($\chi_{\text{spin}}$) \cite{Kittel-2005, Imai-2008}, i.e., $\chi$ = $\chi_{\text{dia}}$ + $\chi_{\text{orb}}$ + $\chi_{\text{spin}}$. Here, for FeSe, $\chi_{\text{dia}}$ = $-$ 6.1 $\times$ 10$^{-5}$ cm$^{3}$/mol from the diamagnetism of the atomic ions, and $\chi_{\text{orb}}$ is $T$-independent unless there is an orbital change such as an orbital ordering.

     Figure 2(a) exhibits the relation of the Knight shift $K (T)$ with the sample susceptibility $\chi (T)$, plotted as $K (T)$ vs $\chi (T)$. At $T$ $\ge$ $T_{\text{nem}}$, $K (T)$ is linear with $\chi (T)$ as expected from above, from which we obtain the value of the constant of the hyperfine coupling to the electron spins at $B_{0}$ $\parallel$ $a\& b$: $A_{\text{spin},\parallel a\&b}$ = 30.4 kOe/$\mu_{B}$, and similarly the corresponding hyperfine coupling constant at $B_{0}$ $\parallel$ $c$: $A_{\text{spin},\parallel c}$ = 32.8 kOe/$\mu_{B}$. As discussed later, the constants ($A_{\text{orb}}$) of the hyperfine coupling to the electron orbitals are also obtained, the values of the spin Knight shift $K_{\text{spin}}$ and orbital shift $K_{\text{orb}}$ are separated, and $\chi_{\text{orb}}$ and $\chi_{\text{spin}}(T)$ are distinguishable, both at $B_{0}$ $\parallel$ $a\&b$ and at $B_{0}$ $\parallel$ $c$.

     Interestingly, at $T$ $<$ $T_{\text{nem}}$, $K(T)$ versus $\chi(T)$ gradually deviates from the high temperature linear relation, as seen in figure 2(a) for both $B_{0}$ $\parallel$ $a\&b$ and $B_{0}$ $\parallel$ $c$. Because $K(T)$ and $\chi(T)$ are fully magnetic in nature, as described by equation (1), this deviation can only be explained by a change in the electron spin susceptibility $\chi_{\text{spin}}$ ($T$) such as that as a result of an AFM order of the electron spins or AFM spin fluctuations, and/or by a change in the electron orbital susceptibility $\chi_{\text{orb}}$ such as that as a result of an ordering of the electron orbitals, as well as associated changes in the hyperfine couplings to the electron spins ($A_{\text{spin}}$) and/or to the electron orbitals ($A_{\text{orb}}$), any of which could lead to a change in $K(T)$ simultaneously. This is also seen by the expression \cite{Slichter-1989, Kotegawa-2008, Imai-2008}
%
\begin{equation}
\begin{aligned}
K(T)&=K_{\text{spin}}(T)+ K_{\text{orb}}=\frac{A_{\text{spin}}}{N_{A}\mu_{B}} \chi_{\text{spin}}(T) + \frac{A_{\text{orb}}}{N_{A}\mu_{B}} \chi_{\text{orb}}  \\
     &=\frac{A_{\text{spin}}}{N_{A}\mu_{B}} \left[ \chi (T) - \chi_{\text{orb}} - \chi_{\text{dia}} \right] + \frac{A_{\text{orb}}}{N_{A}\mu_{B}}\chi_{\text{orb}},
\end{aligned}
\end{equation}
%
where only $K(T)$ and $\chi(T)$ are temperature dependent.
\begin{figure*}
\includegraphics[scale= 0.22]{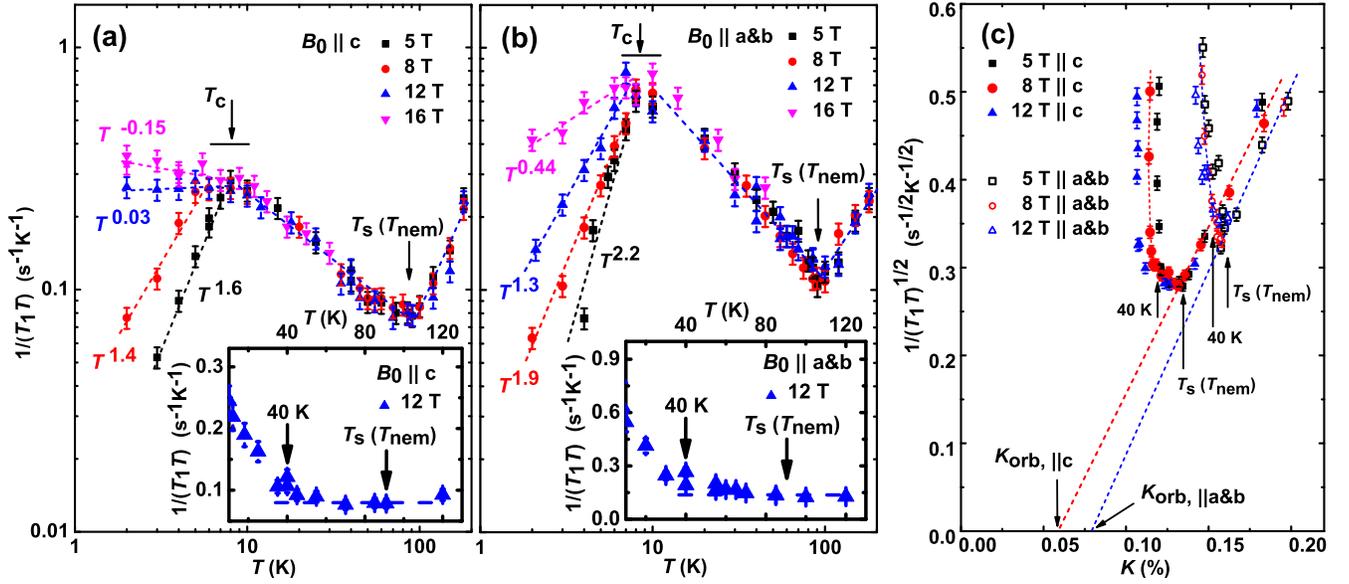}
\caption{Field-dependence of the $^{77}$Se-NMR spin-lattice relaxation in FeSe upon cooling in temperature ($T$). $^{77}$Se $1/T_{1}T$ versus $T$ at (a) $B_{0}$ $\parallel$ $c$ and (b)$B_{0}$ $\parallel$ $a\&b$, respectively. The insets are enlargements in the low $T$ regime. (c) Plot of $\sqrt{1/(T_{1}T)}$ versus Knight shift $K(T)$, where the straight dashed lines are the fits to the Korringa law [Equation (4)].
\label{fig3}}
\end{figure*}

     However, surprisingly, upon further cooling in temperature, the $K(T)$ - $\chi(T)$ plot exhibited in figure 2(a) inset shows that the slope of $K(T)$ versus $\chi(T)$ is $\sim$ 0, both at $B_{0}$ $\parallel$ $a\&b$ and $B_{0}$ $\parallel$ $c$ at $B_{0}$ = 12 T in the temperature range $\sim$ 60 - 10 K, which is a wide range of temperature below $T_{\text{nem}}$ and above the critical temperature $T_{c}$ of superconductivity, i.e., $K_{\text{spin},\parallel a}$ $\approx$ 0, $K_{\text{spin},\parallel b}$ $\approx$ 0, and $K_{\text{spin},\parallel c}$ $\approx$ 0. This is also true for all other fields we applied. That is to say that below $T_{\text{nem}}$ the spin Knight shift $K_{\text{spin}}(T)$ becomes negligible at all directions, i.e., $K$ $\approx$ $K_{\text{orb}}$.

     In other words, the Knight shift $K(T)$ at low temperatures predominantly comes from the contribution of the orbital Knight shift $K_{\text{orb}}$ [Figure 2(b)].

     The reason that $K_{\text{spin}}(T)$ $\approx$ 0 in all directions can be understood by enormous AFM spin fluctuations developed in the same temperature regime, whereas there is no existence of electron spin order, as directly evidenced by our $^{77}$Se-NMR spin-lattice relaxation data (see next), with the consideration of a more general expression of the spin Knight shift as \cite{Slichter-1989, Imai-2008}: $K_{\text{spin}}$ = $\sum_{i}\frac{A^{i}_{\text{spin}}}{N_{A}\mu_{B}}\chi^{i}_{\text{spin}}(T)$. This is the summation of the terms of hyperfine coupling interaction to the individual electron spins (the degree of electron spin polarization is $\propto$ $\chi^{i}_{\text{spin}}$), where each term could be very different from each other due to the AFM spin fluctuations, resulting in a cancelation of them.

     On the other hand, the dramatic increase of the orbital Knight shift $K_{\text{orb}}$ [Figure 2(b)] below $T_{\text{S}}$ ($T_{\text{nem}}$) must be the result of an orbital ordering. To confirm this, we studied the internal field difference ($\Delta B'_{a,b}$) in the $ab$-plane by the measurement of the frequency difference ($\Delta\nu_{a, b}$) of the NMR spectrum peaks (P$_{1}$ and P$_{2}$), as shown in figure 2(c). $\Delta\nu_{a, b}$ reaches $\sim$ 12.5 kHz and 25.0 kHz, or a value of internal field difference $\Delta B'_{a,b}$ $\approx$ 15.6 G and 31.2 G at $B_{0}$ = 8 T and 16 T, respectively, at low temperatures. These values are scalable with $B_{0}$, which is understandable as they are magnetic in nature. Here we have $\Delta B'_{a,b}$ = $\Delta\nu_{a, b}$ /$^{77}\gamma_{Se}$. Since there is no appearance of AFM spin order and the in-plane anisotropy of the paramagnetic spin Knight shift is expected to be negligible (i.e., $K_{\text{spin}, \parallel a}$ $\approx$ $K_{\text{spin}, \parallel b}$), from the Knight shift we have
%
\begin{equation}
\begin{aligned}
\Delta B'_{a,b}&= B_0[(K_{\text{spin}, \parallel a} - K_{\text{spin}, \parallel b}) + (K_{\text{orb}, \parallel a} - K_{\text{orb}, \parallel b}) ]  \\
          & \approx B_{0}(K_{\text{orb}, \parallel a} - K_{\text{orb}, \parallel b}).
\end{aligned}
\end{equation}
%

     Therefore, this indicates that all the data values of $\Delta\nu_{a, b}$ shown in figure 2(c), are essentially completely from the orbital contributions (for convenience, we say all orbital), i.e., the internal field difference in the $ab$-plane is fully determined by the difference of the hyperfine coupling to the Fe-electron orbitals among the $a-$ and $b-$axes. In other words, these data verify that there is an electron orbital ordering occurring at $T$ $\le$ $T_{\text{nem}}$.

     Now we can define a characteristic orbital ordering temperature $T^{+}$ by the intersection of two lines that fit to the data in the transition area as shown in figure 2(c), and we find that $T^{+}$ is linear to $B_{0}$ as:  $T^{+}$ = $T_{\text{nem}}$ $-$ $kB_{0}$, where $k$ = 2.4 $\pm$ 0.1 (K/T) [Figure 4]. We noted that $T^{+}$ indicates the temperature where orbital ordering is fully developed while the orbital ordering starts at $T_{\text{nem}}$ upon cooling. Here $\Delta\nu_{a, b}$ or $\Delta B'_{a,b}$ can be treated as the orbital ordering parameter \cite{Baek-2015}, $\Delta\nu_{a, b}$ $\propto$ $\sqrt{T_{\text{nem}} - T}$ near $T$ = $T_{\text{nem}}$, and as $B_{0}$ $\rightarrow$ 0, $T^{+}$ = $T_{\text{nem}}$.

     Furthermore, in order to investigate the electron spin dynamics and to support the observations in the NMR spectrum and Knight shift, we performed the $^{77}$Se-NMR spin-lattice relaxation measurements as a function of temperature and applied field, as exhibited in figure 3.

     Generally, $1/T_{1}T$ probes the imaginary part of the low-frequency ($\omega$ $\rightarrow$ 0) dynamical susceptibility [$\chi(q, \omega)$] averaged over the momentum ($q$) space as \cite{Slichter-1989, Moriya-1963}: $1/T_{1}T = [3k_{B}/(4\mu^{2}_{B}\hbar^{2})]\sum_{q}A_{q}A_{-a}\chi''(q,\omega)/\omega$, where $A_{q}$ is the hyperfine coupling constant. For conventional Fermi liquid conductors, $\sum_{a}\chi''(q,\omega) = \pi\sum_{k,k'}\delta(E_{k}-E_{k'}-\hbar\omega)(f(E_{k}-E_{k'})$, which gives the Korringa law \cite{Slichter-1989, Moriya-1963}:
\begin{equation}
\begin{aligned}
     1/T_{1}T & = ( \pi/\hbar)A^{2}_{hf}N^{2}(E_{F})k_{B} \\
                & =(4\pi k_{B}/\hbar)(\gamma_{I}/\gamma_{e})^{2}K^{2}_{\text{spin}},
\end{aligned}
\end{equation}
     where $\gamma_{I(e)}$ is the gyromagnetic ratio of nucleus (electron), $N(E_{F})$ is the density of states of electrons at the Fermi energy $E_{F}$, and $f(E)$ is the energy distribution function. For AFM correlated electrons, $\chi(q)$ can have a peak at the AFM wave factor $Q$ = $(\pi, \pi)$, and then $1/T_{1}T$ $\propto$ $\chi(Q)$ with a Curie-Weiss type relation as: $1/T_{1}T$ = $C'/(T-\theta)$, as often seen in cuprate and other Fe-based superconductors \cite{Millis-1990, Aharen-2010, Nakai-2013, Dai-2015}. For AFM fluctuations, the fit parameter $\theta$ $<$ 0, and for large spin fluctuations $C'$ is large.

     Thus, important information can be obtained from the NMR spin-lattice relaxation. First, figures 3(a) and 3(b) show the nematic order/structure phase transition at $T_{\text{nem}}$ = $T_{s}$, which is independent of $B_{0}$. Second, enormous AFM spin fluctuations are developed [a significant deviation from the Korringa law] as seen in the plot of $\sqrt{1/(T_{1}T})$ versus $T$ but they start at $\sim$ 40 K and below only [inset of figures 3(a) and 3(b)], which is far below $T_{\text{nem}}$ = 89 K.  With the fit to the Curie-Weiss relation for 10 K $<$ $T$ $<$ 40 K, we have the values of $\theta$ = $- 4.6$ $(- 21.5)$ K, and $C'$ = 10.0 (7.2) s$^{-1}$ for $B_{0}$ $\parallel a\&b$ ($B_{0}$ $\parallel c)$. Here $\theta$ is comparable while $C'$ is much larger than those of other Fe-based superconductors \cite{Nakai-2008, Ning-2010, Kitagawa-2010, Ma-2011}. Third, the AFM spin fluctuations drop significantly at $T$ $<$ $T_{c}$ due to diamagnetism associated with the pairing symmetry of the electron spins, and $1/T_{1}$ $\propto$ $T^{\alpha}$, where $\alpha$ $\approx$ 3 in low fields, consistent with a line-node gap behavior of a $d$-wave superconductor, agreeing with reports on various Fe-based superconductors \cite{Kotegawa-2008, Nakai-2008, Ning-2008, Imai-2009}.

     Figure 3(c) exhibits the plot of $\sqrt{1/(T_{1}T})$ versus $K(T)$ with $T$ as an implicit parameter, with the consideration that the Korringa law [Equation (4)] can also be expressed as $\sqrt{1/(T_{1}T})$ = $\sqrt{C}K_{\text{spin}}(T)$ = $\sqrt{C}$ $[K(T)- K_{\text{orb}}]$ for a Fermi liquid. Here $C$ = ($4\pi k_{B}/\hbar)(\gamma_{I}/\gamma_{e})^{2}$ for free electrons \cite{Slichter-1989, Moriya-1963}. Apparently, figure 3(c) shows a linear relation above $T_{\text{nem}}$, and thus it gives values of $K_{\text{orb}}$ $\approx$ 0.06$\%$ (0.08$\%$) for $B_{0}$ $\parallel$ $c$ ($B_{0}$ $\parallel$ $a\&b$) by the intercepts along the $K(T)$ axis, which have been used to separate $K_{\text{spin}}$ and $K_{\text{orb}}$ in the tetragonal phase [shown in figure 2(b)] and to extrapolate the values of $A_{\text{orb}}$, $A_{\text{spin}}$, $\chi_{\text{orb}}$, and $\chi_{s}(T)$ combining with the $K(T) - \chi(T)$ relation [Figure 3(c)]. Similarly, the slope also gives an experimental value of $C$ $\approx$ 1.5 $\times$ 10$^{5}$ (1.8 $\times$ 10$^{5}$) K$^{-1}$s$^{-1}$ for $B_{0}$ $\parallel$ $c$ ($B_{0}$ $\parallel$ $a\&b$), which matches well with the theoretical value of $C$ = 1.46 $\times$ 10$^{5}$ K$^{-1}$s$^{-1}$ for non-interacting/free electrons in FeSe. Thus, these data verify that the electrons at $T$ $>$ $T_{\text{nem}}$ in FeSe are not strongly correlated.

     Moreover, below $T_{\text{nem}}$ in the range 40 K$<$$T$$\leq$$T_{\text{nem}}$, $1/T_{1}T$ continues to show a free-electron behavior (Korringa law), where $1/T_{1}T$ is a constant as exhibited in figures 3(a)-3(b) insets, i.e., there is essentially no AFM spin fluctuations at the $T_{\text{nem}}$ regime over a wide range of temperature. Therefore, AFM spin fluctuations can be excluded from the driving mechanism of the nematic order.

     Figure 4 shows temperature - field phase diagram we obtained. First, as described in detail above, the applied magnetic field decreases the characteristic orbital ordering temperature $T^{+}$ rather sensitively. Second, the structural phase transition temperature $T_{\text{s}}$ and the nematic ordering temperature $T_{\text{nem}}$ are not affected by the applied field ($T_{\text{nem}}$ = $T_{\text{s}}$ = 89 K). Third, stripe-type AFM order is absent at all the applied fields. Here the values of $T_{c}$ were determined by our resonance frequency method \cite{cao-2018}.

     We would like to point out that, among the two groups of the 3$d$ Fe $t_{2g}$ ($d_{xy}$, $d_{yz}$, $d_{xz}$) and $e_{g}$ ($d_{x^2-y^2}$, $d_{3z^2-r^2}$) orbitals (totally five orbitals), $d_{xy}$, $d_{x^2-y^2}$, and $d_{3z^2-r^2}$ are rotationally symmetric in the $xy$-plane. Thus, the only two candidates related to the tetragonal-orthorhombic degeneracy (lattice symmetry breaking) are the $d_{yz}$ and $d_{xz}$ orbitals. And, what an NMR spectrum directly measures is the local field distribution and local field values parallel to the externally applied magnetic field at the nucleus at the atomic scale. Therefore, because of the unique symmetry of the the $d_{yz}$ and $d_{xz}$ orbitals in the lattice axis $c$-direction the ordering of these $d_{yz}$ and $d_{xz}$ orbitals is not able to cause any splitting of the $^{77}$Se-NMR spectra at $B_{0}$ $\parallel$ $c$ in FeSe.
\begin{figure}
\includegraphics[scale= 0.12]{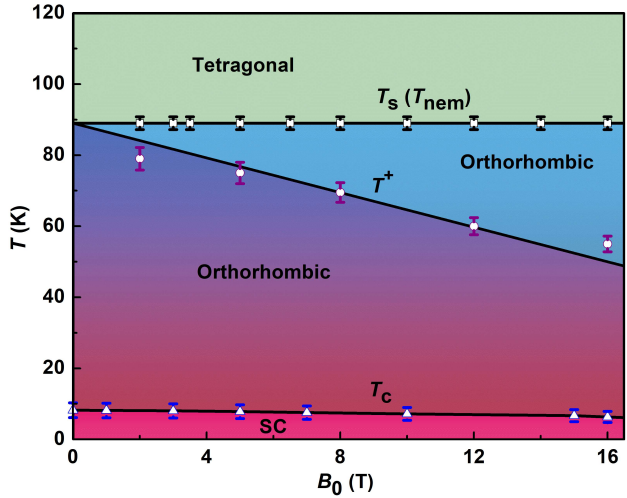}
\caption{The temperature-field ($T$ - $B_{0}$) phase diagram of FeSe. The obtained phase diagram of FeSe in applied magnetic field $B_{0}$ with temperatures $T_{\text{s}}$, $T_{\text{nem}}$, $T^{+}$, and $T_{\text{c}}$ (see text for definition). The solid lines are guides to the eyes.
\label{fig4}}
\end{figure}

     Finally, we discuss the field effect on the characteristic temperatures. That the values of $T_{\text{s}}$ ($T_{\text{nem}}$) are not affected by the directions or magnitude of the applied field could be explained by the weak anisotropy character of the paramagnetic Fe-spins in the high symmetry tetragonal lattice. That $T^{+}$ is linearly proportional to $B_{0}$ could be understood due to its full magnetic character that involves electron orbital moments, while the reason for the decrease of the value of $T^{+}$ with $B_{0}$ is not clear.

     Whereas there is no appearance of long-range AFM order of the electron spins, we note that there were reports about short-range stripe magnetic order and possible spin-orbital coupling \cite{Ma-2017, Day-2018, He-2018} developing in the nematic phase. From our data of the temperature- and field-dependence of  $^{77}$Se-NMR spin-lattice relaxation, 1/$T_{1}T$ versus $T$, we clearly see that giant AFM fluctuations gradually develop starting at $\sim$ 40 K and below, which is far below the nematic order temperature $T_{\text{nem}}$. And above $T_{\text{nem}}$ the data of 1/$T_{1}T$ well follows the Korringa law, which is a direct evidence of trivial electron correlations. These results leave the orbital ordering unequivocally as the dominant driving force of the nematic order.
\section*{4. Conclusions}
     In summary, we report direct observation of orbital ordering which is demonstrated by the splitting of the $^{77}$Se-NMR spectrum and Knight shift in single crystals of iron-based superconductor FeSe. As illustrated by the field-dependence of the $^{77}$Se-NMR spin-lattice relaxation, stripe-type AFM order is absent, whereas giant AFM spin fluctuations developed starting at temperatures far below the electronic nematic order temperature $T_{\text{nem}}$, thus both of which can be unambiguously excluded from the origin of the nematic order. These discoveries provide direct evidence of orbital-driven nematic order in FeSe. Our results also help to the understanding of the strong interplay between structure, magnetism and superconductivity in Fe-based superconductors as well as other unconventional superconductors.
%
%
%
%
%
%
\section*{Acknowledgements}
     Work at YZU was supported by National Science Foundation of China (NSFC) (Grants $\#$ 61474096 and 1804291) and NSF of Jiangsu (Grants $\#$ BK20180889 and BK20180890), and at CAS by NSFC (Grants $\#$ 51477167 and 41527802). D.A.C. thanks supports by the program 211 of the Russian Federation Government (RFG), agreement 02.A03.21.0006 and by the RFG Program of Competitive Growth of KFU. A.N.V. thanks supports by Russian Foundation for Basic Research Grant $\#$ 17-29-10007, by the Ministry of Education and Science of the RFG in the framework of ICP of NUST MISiS (Grant $\#$ K2-2017-084), and by Act 211 of RFG, agreements 02.A03.21.0004, 02.A03.21.0006, and 02.A03.21.0011.
%
%
%
\section*{Keywords}
     nuclear magnetic resonance (NMR), NMR spin-lattice relaxation, nematic order, orbital ordering, Fe-based superconductor.

%
%
%
%
%

%

\section*{References}

\end{document}